# Spin transfer torque devices utilizing the giant spin Hall effect of tungsten

Chi-Feng Pai,[1,a] Luqiao Liu,[1] Y. Li,[1] H. W. Tseng,[1] D. C. Ralph,[1,2] and R. A. Buhrman[1]

[1]Cornell University, Ithaca, New York 14853, USA

[2]Kavli Institute at Cornell, Ithaca, New York, 14853, USA

We report a giant spin Hall effect (SHE) in $\beta$-W thin films. Using spin torque induced ferromagnetic resonance with a $\beta$-W/CoFeB bilayer microstrip we determine the spin Hall angle to be $\left|\theta_{SH}^{\beta\text{-W}}\right| = 0.30 \pm 0.02$, large enough for an in-plane current to efficiently reverse the orientation of an in-plane magnetized CoFeB free layer of a nanoscale magnetic tunnel junction adjacent to a thin $\beta$-W layer. From switching data obtained with such 3-terminal devices we independently determine $\left|\theta_{SH}^{\beta\text{-W}}\right| = 0.33 \pm 0.06$. We also report variation of the spin Hall switching efficiency with W layers of different resistivities and hence of variable ($\alpha$ and $\beta$) phase composition.

[a] Author to whom correspondence should be addressed. Electronic mail: cp389@cornell.edu

In the spin Hall effect (SHE)[1-3] the application of a charge current through a nonmagnetic material results in the generation of a transverse spin current due to the spin-orbit interaction. This transverse spin current can be described by $J_s = \theta_{SH}(\sigma \times J_e)$, where $\hbar J_S / 2e$ is the spin current density, $J_e$ the charge current density, $\sigma$ the spin polarization unit vector and $|\theta_{SH}| = |J_s / J_e|$ is the material-dependent spin Hall angle. Previous studies have demonstrated that the magnitudes of the spin Hall angles in the 5d elements Pt[4-9] and high resistivity $\beta$-Ta[10] can be relatively large, $\theta_{SH}^{Pt} = 0.07$ and $\theta_{SH}^{\beta\text{-Ta}} \approx -0.15$. The resulting spin currents are sufficiently strong to be of interest for magnetization manipulation via the spin transfer torque mechanism[11-13], opening the possibility of an efficient new way to implement spintronics technologies.

The particularly large spin Hall angle found in high resistivity $\beta$-Ta motivates the study of other transition metals with large atomic numbers that can also be formed in a high resistivity phase. Research has shown that very thin tungsten films are routinely formed in a $\beta$-W phase with resistivity $\rho_{\beta\text{-W}} = 100\text{-}300\ \mu\Omega\cdot\text{cm}$ that has been determined to have the A15 crystal structure associated with very strong electron-phonon scattering, while somewhat thicker and/or annealed W films tend to be of a mixed $\alpha$ and $\beta$ composition, and still thicker films are typically purely $\alpha$-W with $\rho_{\alpha\text{-W}} \leq 25\ \mu\Omega\cdot\text{cm}$ and a bcc crystal structure[14-17]. Among the 5d transition metals, W has been predicted to have the largest spin Hall conductivity in terms of magnitude while in a highly resistive state[18], although to the authors' best knowledge the SHE in W has not yet been studied experimentally. Here we report, from spin torque ferromagnetic resonance (ST-FMR) measurements on $\beta$-W/CoFeB bilayer microstrips, the existence of a very large spin Hall angle in $\beta$-W, $\left|\theta_{SH}^{\beta\text{-W}}\right| \approx 0.3$, twice as large as the previous record high value from $\beta$-Ta[10]. We independently confirm this result by demonstrating efficient magnetic switching driven by the spin Hall effect spin torque (SHE-ST) in 3-terminal nanoscale magnetic tunnel junction (MTJ) devices with a CoFeB free layer and a $\beta$-W base. Finally, by performing SHE-ST switching measurements of MTJs formed on W thin films with compositions ranging from the high resistivity $\beta$ phase to low resistivity $\alpha$ phase, we show a direct correlation between the W resistivity (phase composition) and $\left|\theta_{SH}\right|$.

The W films studied here were produced by dc-magnetron sputtering onto oxidized Si substrates at a base pressure of < 2 x 10$^{-8}$ Torr. The deposition rate was ≤ 0.02 nm/sec to minimize film heating (essential for obtaining the $\beta$-W phase). Previous work has shown that unlike Ta, the resistivity of sputtered W films depends on deposition conditions and also has a strong thickness ($t$) dependence[16,17]. This is consistent with our measurements of resistivity versus $t$ shown in Fig. 1(a). Our W films exhibit a resistivity as high as $300\ \mu\Omega\cdot\text{cm}$ for our thinnest layers ($t$ = 4 nm), a signature of the $\beta$-W phase. The resistivity decreases with increasing $t$ until for $t \geq 8$ nm it saturates at $\approx 20\ \mu\Omega\cdot\text{cm}$, in the range expected for a purely $\alpha$-W phase.

With X-ray diffraction we further verified the existence, within measurement accuracy, of only the α-W phase (bcc) in a W(8) film (thickness in nanometers), and a mixture of α-W and β-W phases (bcc+A15) in a W(6) sample [Fig. 1(b)], in agreement with previous studies[15,16] and consistent with the resistivity variation. The W(4) samples provided too little X-ray signal for us to make an unambiguous phase determination. Previous work has shown that thin W films in the β phase can be converted partially into α-W by thermal annealing[14,16]. We found that thermal processing to 170 °C of our multilayers with W base layer thicknesses ≤ 6 nm did tend to reduce the resistivity by about 20%, but the W resistivity was then stable for subsequent thermal annealing as high as 350 °C.

We determined the spin Hall angle of β-W first by performing ST-FMR measurements on micron-sized W(6)/Co$_{40}$Fe$_{40}$B$_{20}$(5) bilayer structures. The in-plane anisotropy of the magnetic moment of the CoFeB(5) was verified by SQUID measurements on as-grown films. The effective demagnetization field $\mu_0 M_{eff} = 1.3 \pm 0.05$ T and the saturated magnetization $M_s = (1.2 \pm 0.01) \times 10^6$ A/m of the CoFeB layer were determined independently by magnetometry. In these bilayers the CoFeB resistivity was $\rho_{CoFeB} \approx 168\ \mu\Omega\cdot cm$ and the W resistivity was 170 $\mu\Omega\cdot cm$, indicative of a β-W rich composition. In comparison, the resistivity of the 6 nm uncapped W layer used for the X-ray diffraction study was 90 $\mu\Omega\cdot cm$. The bilayers were patterned into 10 μm wide bars similar to the samples used in previous ST-FMR studies of Pt[7] and β-Ta-based[10] systems.

For the ST-FMR measurement, an in-plane microwave frequency (RF) current $I_{RF}$ is sent through the bilayer samples as shown in Fig. 2(a), and that portion of the oscillating current that flows within the W layer generates both an out-of-plane Oersted field torque and an in-plane SHE-ST acting upon the magnetic moment in the adjacent CoFeB layer. When the oscillating current frequency $f$ and the magnitude of the applied external magnetic field $\mu_0 H$ satisfy the ferromagnetic resonance condition, which can be described by the Kittel formula $f = (\gamma/2\pi)\mu_0\sqrt{H(H+M_{eff})}$ ($\gamma$ is the gyromagnetic ratio), the precession of the magnetic moment results in a DC voltage $V_{mix}$ due to the mixing of $I_{RF}$ and the oscillating anisotropic magnetoresistance of the CoFeB layer. The ST-FMR $V_{mix}$ signal as a function of external magnetic field $\mu_0 H$ consists of a simple sum of symmetric and anti-symmetric Lorentzian peaks, originating respectively from the SHE-ST and the Oersted field torque. The linewidth $\Delta$ of the peaks is proportional to the effective damping constant $\alpha'$ of the CoFeB magnetic moment by $\Delta = (2\pi f/\gamma)\alpha'$ and can be increased or decreased by the SHE-ST from a DC charge current $I_{DC}$, depending on its polarity and hence the direction of the spin polarization injected into the CoFeB layer. Quantitatively, the effective damping constant $\alpha'$ as a function of the DC charge current density $J_e$ should have the form[19,20]

$$\alpha' = \alpha_0 + \frac{\sin\phi}{\mu_0 M_s t\left(H + M_{eff}/2\right)} \frac{\hbar}{2e} \theta_{SH} J_e, \quad (1)$$

where $\alpha_0$ is the damping constant of the CoFeB layer when $I_{DC}=0$ and $\phi$ is the angle between $I_{DC}$ and $\mu_0 H$. The dependence of the effective damping on $J_e$ is shown in Fig. 2(b) for both $\phi = 45°$ (positive field) and $\phi = -135°$ (negative field), as measured with $f$ = 9 GHz. $J_e$ in the W layer was determined from $I_{DC}$ by using the measured resistivities of the W and CoFeB layers. By fitting these data to Eq. (1), we obtained $\alpha_0 = 0.0122 \pm 0.0004$ and $\left|\theta_{SH}^{\beta\text{-W}}\right| = 0.30 \pm 0.02$ with the sign of $\theta_{SH}^{\beta\text{-W}}$ being negative (the same as for Ta and opposite to Pt), consistent with theoretical predictions[18,21]. This should represent a lower bound on the true bulk value of $\left|\theta_{SH}^{\beta\text{-W}}\right|$, since the measured value can be suppressed if the W film thickness is less than or comparable to its spin diffusion length.

To independently confirm this exceptionally large $\theta_{SH}^{\beta\text{-W}}$ and to demonstrate the efficacy of the SHE in $\beta$-W for spintronics applications, we fabricated 3-terminal devices consisting of a W microstrip with a CoFeB/MgO/CoFeB nano-pillar magnetic tunnel junction (MTJ) formed on top [Fig. 3(a)]. To make these devices we first sputter-deposited on a thermally-oxidized Si substrate a multilayer with the structure W(5.2)/Co$_{40}$Fe$_{40}$B$_{20}$(2)/MgO(1)/ Co$_{40}$Fe$_{40}$B$_{20}$(4)/Ta(4)/Ru(5) and then patterned it by two steps of electron-beam lithography and ion-milling. When milling the nanopillar MTJ, we over-etched into the W layer by nominally 2 nm to ensure that the bottom

CoFeB free layer was fully patterned. After deposition of a SiO$_2$ insulating layer and top contacts, post-fabrication annealing was applied at 250$^\circ$C for one hour to enhance the tunneling magnetoresistance (TMR) of the MTJ. In the device whose behavior is discussed here in detail the MTJ pillar had a cross section ~ 80 nm × 300 nm with its long axis perpendicular to the current direction in the W channel direction. The bottom W channel was 1.2 μm in width and 6 μm in length with a channel resistance ≈ 4 kΩ, which corresponds to $\rho_W \approx 260 \pm 40\ \mu\Omega \cdot \text{cm}$ [22], indicative of predominantly the $\beta$-W phase. We observed consistent behavior in three devices with nano-pillar dimensions ranging from 60 nm × 200 nm to 100 nm × 350 nm.

The magnetic-field-induced switching behavior of the device is shown in Fig. 3(b). We applied an in-plane external magnetic field $B_{ext}$ parallel to the easy axis of the MTJ and measured the differential resistance (*dV/dI*) by a lock-in method with a sensing current $I_{ac} = 0.1\ \mu\text{A}$. The TMR minor loop of the device in Fig. 3(b) indicates a parallel state MTJ resistance $R_P \approx 12.6\ \text{k}\Omega$, a coercive field $B_C \approx 5$ mT and (after subtracting the 2 kΩ lead resistance) a TMR ≈ 51%.

Magnetization switching driven by an in-plane DC current within the W channel is shown in Fig. 4(a). Here we applied an in-plane external magnetic field of -15 mT to cancel the dipole field from the top fixed layer. The free layer of the MTJ can be switched reproducibly between the antiparallel (AP) state and parallel (P) state of the MTJ at the critical currents $I_{DC} \approx \pm 0.32$ mA. The polarity of the current-induced switching is consistent with our previous results for $\beta$-Ta-based devices[10], which again corresponds to $\theta_{SH} < 0$. We also measured the dependence of the critical switching currents on the ramp rate of the current sweep [Fig. 4(b)]. By fitting these data to the thermally-assisted spin torque switching model[23], we found that the zero-thermal-fluctuation critical currents for AP to P and P to AP switching are the same to within measurement uncertainty, $|I_{c0}| = 0.95 \pm 0.03$ mA, with an energy barrier between the two states $U = 40.4 \pm 0.3\ k_B T$. Considering the geometry of the device, the critical current density is $J_{c0} = (1.8 \pm 0.3) \times 10^{11}$ A/m$^2$.

The zero-thermal-fluctuation critical current density $J_{c0}$ for SHE-ST switching of an in-plane polarized ferromagnet by the anti-damping mechanism can be written approximately as[24,25]

$$J_{c0} \approx \frac{2e}{\hbar} \mu_0 M_s t \alpha_0 \left( H_c + \frac{M_{eff}}{2} \right) / \theta_{SH}, \qquad (2)$$

where $H_c$ in the coercive field of the free layer. We measured $M_s = (1.18 \pm 0.02) \times 10^6$ A/m by magnetometry on an unpatterned W(5.2)/CoFeB(2)/MgO(1)/Ru(5) test sample. After patterning this test sample, we determined $\alpha_0 = 0.0186 \pm 0.0007$ for the 2 nm CoFeB layer using ST-FMR measurements[7] and $\mu_0 M_{eff} = 0.89 \pm 0.03$ T by anomalous Hall resistance measurements as a function of an out-of-plane magnetic field. All the samples used for determining $M_s$, $\alpha_0$, and $M_{eff}$ were annealed at 250°C for one hour before measurements, the same annealing procedure used for the full 3-terminal MTJ samples. Using these parameters with Eq. (2), we obtained $|\theta_{SH}^{\beta\text{-W}}| = 0.33 \pm 0.06$ for the 5.2 nm W base layer, thereby providing independent confirmation of the result obtained from the ST-FMR linewidth-vs.-$I_{DC}$ measurements (Eq. (1)).

To explore the spin Hall efficiencies in different phases of W, we fabricated 3-terminal devices with two additional W channel thicknesses: 6.2 nm and 15 nm. The width of the bottom W channel and the size of the MTJ nano-pillar were held the same in all these devices, within the precision of the fabrication process. The results are summarized in Table. I. For the W(6.2) 3-terminal device, the W channel resistivity was measured to be $\approx 80\ \mu\Omega\cdot\text{cm}$, approximately 1/3 the resistivity of the 5.2 nm W device, indicative of a mixed α-W and β-W composition. The switching behavior of the W(6.2) 3-terminal device was similar to the W(5.2) device but with a larger critical current density ($J_{c0} = (3.2 \pm 0.2) \times 10^{11}$ A/m$^2$), corresponding to a smaller spin Hall angle, $|\theta_{SH}| = 0.18 \pm 0.02$. This is still greater than our result for β-Ta[10], where $|\theta_{SH}^{\beta\text{-Ta}}| \approx 0.15$ for films with $\rho_{\beta\text{-Ta}} \approx 190\ \mu\Omega\cdot\text{cm}$. In contrast, for the W(15) 3-terminal device, which had a W channel resistivity of $21\ \mu\Omega\cdot\text{cm}$ indicative of a pure α-W phase, we observed no clear signature of SHE-ST switching up to $J_e \approx 6 \times 10^{11}$ A/m$^2$, where current-induced Oersted field switching began (with polarity opposite to the SHE switching). Using that the SHE torque is weaker than the torque due to the Oersted field in the W(15) sample, we can estimate an upper bound for the spin Hall angle in the W(15) 3-terminal device to be $|\theta_{SH}| < 0.07$. This strong variation of $|\theta_{SH}|$ with respect to the W thickness and resistivity is consistent with the results from ST-FMR measurements performed on W/NiFe bilayers with various W thicknesses, which will be reported elsewhere.

As previously discussed[10], the ability of an in-plane charge current within a normal metal layer having a large $|\theta_{SH}|$ to exert a strong spin torque on an adjacent magnetic layer in a 3-terminal device configuration is very attractive for high-performance, high-reliability ST magnetic memory and spin logic applications, in comparison to conventional two terminal ST devices. The very large values of $|\theta_{SH}|$ in W thin films of both the $\beta$ phase ($|\theta_{SH}| \approx 0.3$) and the lower resistivity, mixed $\alpha/\beta$ phase ($|\theta_{SH}| \approx 0.18$) appear to make W particularly well suited for such applications. The ability to obtain $|\theta_{SH}^{W}| > |\theta_{SH}^{Ta}|$ in W films with resistivity < 50% that of $\beta$-Ta could be advantageous for optimizing the write impedance of the 3-terminal SHE device and minimizing the writing energy. The strong correlation between the spin Hall strength and the W film resistivity (and A15 composition) also provides guidance towards developing a more complete understanding of the mechanism of this giant spin Hall effect and towards discovering useful materials with even larger $|\theta_{SH}|$.

**Acknowledgements**

We thank Dr. Maura S. Weathers for assistance in performing the X-ray diffraction on W samples. This work was supported in part by DARPA, ARO, ONR, NSF/MRSEC (DMR-1120296) through the Cornell Center for Materials Research (CCMR), and NSF/NSEC through the Cornell Center for Nanoscale Systems. We also acknowledge support from the NSF through use of the Cornell Nanofabrication Facility/NNIN and the CCMR facilities.

which is ±0.5 nm.

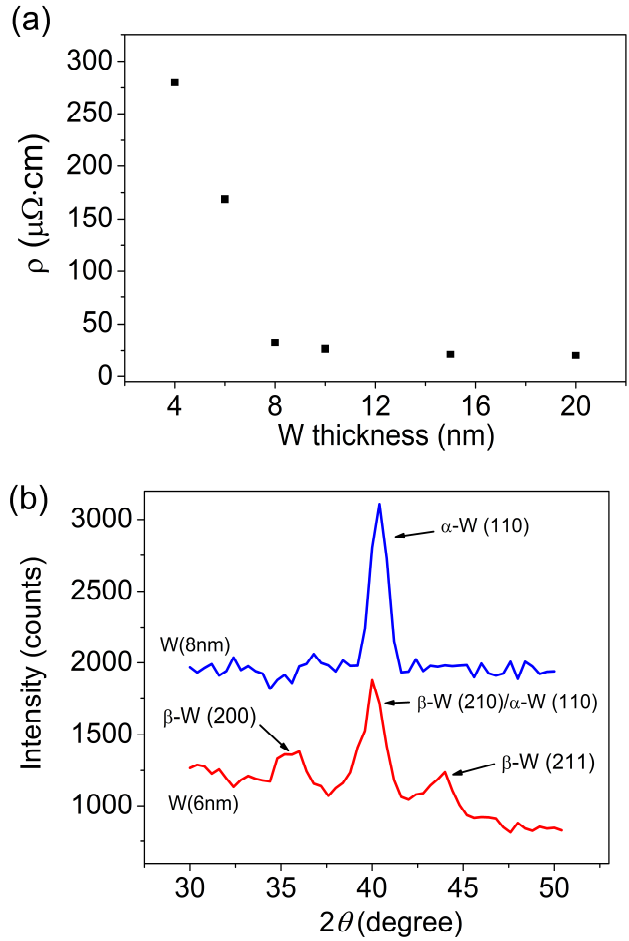

FIG. 1. (Color online) (a) Resistivity of our sputtered W films as a function of thickness. The films are capped by CoFeB and the CoFeB conductance is subtracted. (b) X-ray diffraction patterns for sputtered W(8 nm) (upper blue trace line) and W(6 nm) (lower red trace line) films (without a CoFeB cap). The arrows indicate the identifications of the Bragg peaks and the corresponding diffraction planes.

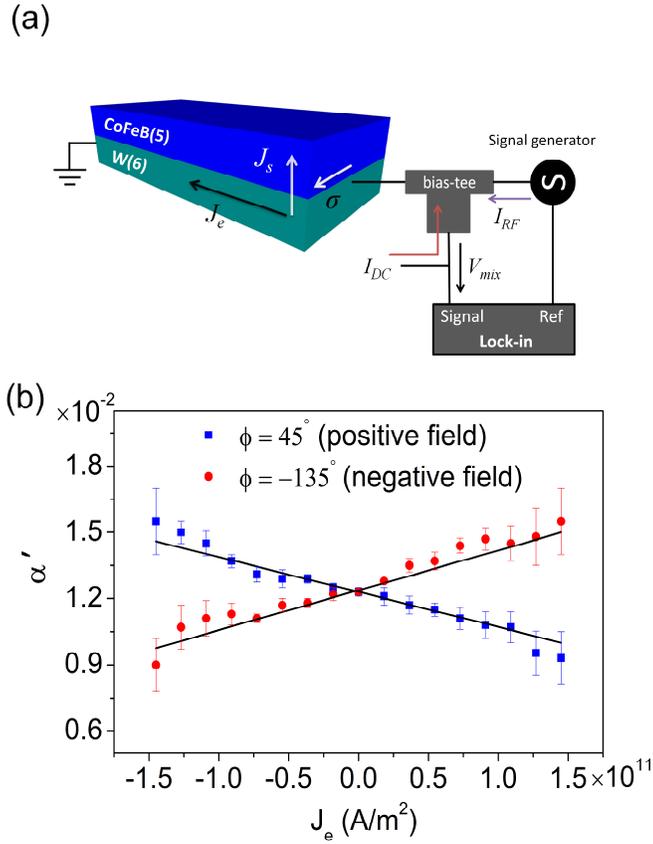

FIG. 2. (Color online) (a) Schematic illustration of the ST-FMR device and the circuit layout for measurements of the resonance linewidth versus DC current. (b) The change of the effective damping constant as a function of DC current charge density $J_e$ for two angles of applied magnetic field. The RF frequency $f$ = 9 GHz. The solid lines represent fits to Eq. (1).

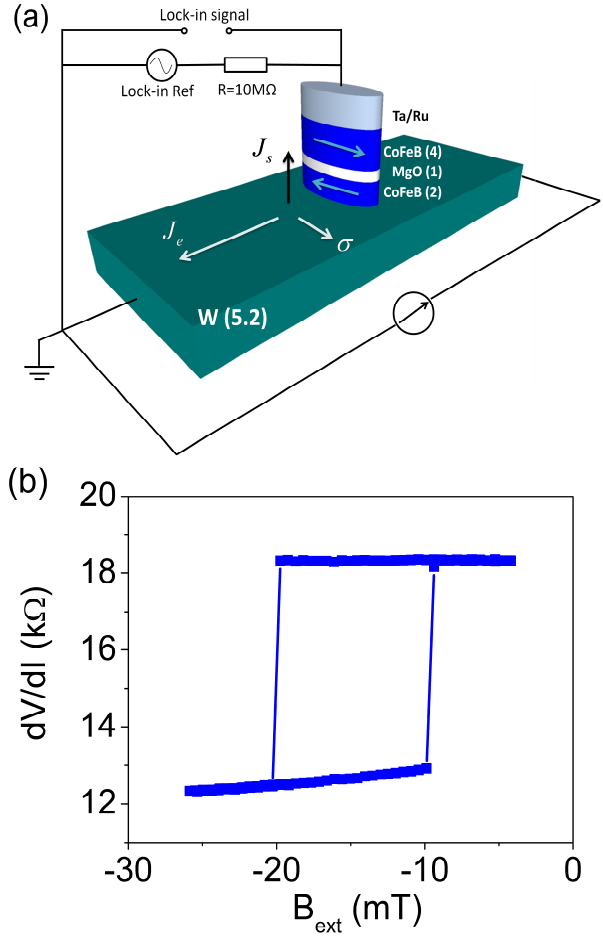

FIG. 3. (Color online) (a) Schematic illustration of the $\beta$-W 3-terminal device and the measurement circuit layout. (b) Differential resistance ($dV/dI$) minor loop for the MTJ without any DC current along the $\beta$-W bottom channel. The measured resistance includes a 2 k$\Omega$ lead resistance from the W channel that is not subtracted here. The external magnetic field $B_{ext}$ was applied along the long axis of the nano-pillar MTJ.

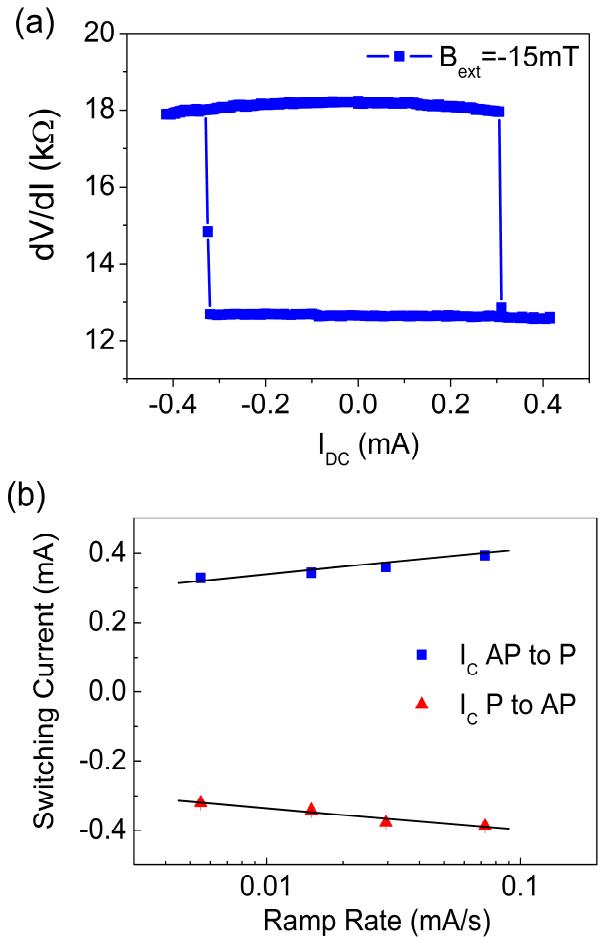

FIG. 4. (Color online) (a) Differential resistance of the MTJ as a function of the DC bias current $I_{DC}$, exhibiting magnetization switching by the spin Hall torque. An in-plane external magnetic field $B_{ext} = -15$ mT was applied to cancel the dipole field from the fixed layer. (b) Switching currents as a function of the current ramp rate. The blue squares represent switching currents from the AP to P state and the red triangles represent switching currents from P to AP. Solid lines are fits to a thermally-assisted spin torque switching model.

TABLE I. The spin Hall angle of W as determined from the critical currents for magnetization switching in 3-terminal ST devices.

| W thickness (nm) | W resistivity ($\mu\Omega\cdot$cm) | Phase | Spin Hall angle $|\theta_{SH}|$ |
|---|---|---|---|
| 5.2 | 260 | $\beta$ | 0.33±0.06 |
| 6.2 | 80 | $\alpha+\beta$ | 0.18±0.02 |
| 15 | 21 | $\alpha$ | < 0.07 |